\documentclass[manuscript=article]{achemso} 

\usepackage[version=3]{mhchem} 

\usepackage{color} 
\usepackage{subcaption}  
\usepackage{braket} 
\usepackage{chemfig} 

\usepackage{hyperref}

\author{Marlon F. Jost}
\affiliation[Northeastern University]
{Department of Chemistry and Chemical Biology, Northeastern University, Boston, Massachusetts 02115, United States}
\alsoaffiliation[Northeastern University]
{Department of Physics, Northeastern University, Boston, Massachusetts 02115, United States}

\author{Sijia S. Dong}
\affiliation[Northeastern University]
{Department of Chemistry and Chemical Biology, Northeastern University, Boston, Massachusetts 02115, United States}
\alsoaffiliation[Northeastern University]
{Department of Physics, Northeastern University, Boston, Massachusetts 02115, United States}
\alsoaffiliation[Northeastern University]
{Department of Chemical Engineering, Northeastern University, Boston, Massachusetts 02115, United States}

\email{s.dong@northeastern.edu}

\title
  {Excited-State Quantum Chemistry on Qumode-Based Processors via Variational Quantum Deflation}

\begin{document}

\begin{abstract}

Variational quantum algorithms on bosonic quantum processors are an emerging paradigm for quantum chemistry calculations, exploiting the natural alignment between molecular structure and harmonic oscillator-based hardware. We introduce the qumode-based variational quantum deflation framework (QumVQD) for finding both electronic and vibrational excited state energies on qumode-based architectures. We validate the approach through electronic structure calculations on \ce{H2} and linear \ce{H4}, where we introduce Hamming-weight filtering of the Fock basis to enforce particle number conservation and eliminate spurious eigenstates by reducing the required Hilbert space, which reduces the required number of qumodes in turn. We achieve agreement with full configuration interaction (FCI) using the STO-3G basis set within the chemical accuracy threshold at most points along the potential energy surfaces. Extending to the vibrational structure, we combine QumVQD with an existing Hamiltonian fragmentation approach based on Cartan subalgebra, allowing us to compute the vibrational eigenenergies of \ce{CO2} and \ce{H2S} to spectroscopic accuracy with per-fragment circuits that scale as $\mathcal{O}(N)$ in single-qumode gates and $\mathcal{O}(N^2)$ in beam-splitter gates for $N$ qumodes. For the case of \ce{CO2}, we get total gate counts more than an order of magnitude smaller than those reported for qubit-based vibrational algorithms at this system size. These results demonstrate that bosonic quantum devices are a viable platform for excited-state quantum chemistry, particularly for vibrational problems where qubit-based methods incur substantial boson-to-qubit mapping overhead.

\end{abstract}

\section{Introduction}

The exponential cost of representing many-body wavefunctions on classical hardware makes quantum chemistry a natural target for quantum computing~\cite{lanyon_towards_2010, cao_quantum_2019, mcardle_quantum_2020}, particularly in high-accuracy or strongly entangled regimes where an explicit wavefunction is unavoidable~\cite{cao_quantum_2019}. To date, most quantum computing research has focused on qubit-based architectures~\cite{wang_qudits_2020}, where quantum information is encoded in two-level systems. While significant progress has been made, current qubit-based quantum computers remain limited by modest qubit counts, restricted gate counts, short coherence times, and high error rates~\cite{preskill_quantum_2018}, constraining their ability to tackle practical quantum chemistry problems.

The recent emergence of highly controllable bosonic quantum devices offers a potentially paradigm-shifting alternative quantum hardware platform for computational chemistry. Unlike qubit-based platforms, the fundamental building block of these bosonic devices is a harmonic oscillator, known as a qumode. Whereas a qubit's state resides in a two-dimensional Hilbert space, a qumode's state lives in an infinite-dimensional Hilbert space that can be represented in either the discrete basis of harmonic oscillator stationary states or a continuous basis of position or momentum operator eigenfunctions. This expanded dimensionality presents several potential advantages over qubits. First, the larger Hilbert space per physical mode is likely to reduce circuit depth requirements for quantum algorithms~\cite{wang_qudits_2020}. Second, since the simulation of vibrational dynamics and spectroscopy is naturally formulated in terms of bosonic operators, working directly in the oscillator space avoids a boson-to-qubit mapping overhead as required with qubit-based devices~\cite{malpathak_simulating_2025}.

Among the various hardware implementations for bosonic quantum computing, circuit quantum electrodynamics (cQED) platforms have emerged as particularly promising candidates~\cite{copetudo_shaping_2024}. These cQED processors utilize microwave resonators as qumodes coupled to superconducting transmon qubits, enabling precise control over the bosonic modes through native gate operations tailored to the harmonic oscillator structure. This architecture allows for potentially more compact state representations compared to qubit encodings, as the expanded Hilbert space dimensionality per physical mode can reduce circuit depth requirements for quantum algorithms~\cite{dutta_simulating_2025, dutta_simulating_2024}.

Qudits and qumodes have been studied for a range of applications~\cite{wang_qudits_2020, dutta_simulating_2024}, including molecular vibronic spectra calculations via boson sampling~\cite{huh_boson_2015}, analog quantum simulation~\cite{macdonell_predicting_2023}, and digital techniques~\cite{wang_efficient_2020}, simulation of chemical quantum dynamics~\cite{lyu_mapping_2023, vu_computational_2025}, ground-state electronic structure calculations of bond dissociation on bosonic quantum devices~\cite{dutta_simulating_2025}, quantum phase estimation with higher-dimensional systems~\cite{parasa_quantum_2011, ye_quantum_2011}, and ansatz design for polaritonic chemistry across qubit, qudit, and hybrid qubit-qumode architectures~\cite{1l5j-dfh4}. Particularly relevant here is the framework for implementing the variational quantum eigensolver (VQE) algorithm on qumode-based devices established by Dutta et al.~\cite{dutta_simulating_2025}, which enables further extensions of VQE to problems beyond ground-state electronic structure.

For problems requiring information beyond the ground state, such as molecular spectroscopy or photochemistry, extensions of VQE that can access excited states become essential. Several such extensions have been developed for qubit-based algorithms, including orthogonality-penalty approaches like variational quantum deflation (VQD)~\cite{higgott_variational_2019}, subspace-search methods such as SSVQE~\cite{PhysRevResearch.1.033062}, and the quantum equation-of-motion (qEOM) approach~\cite{PhysRevResearch.2.043140}. Among these, VQD systematically targets excited states one at a time by augmenting the VQE cost function with orthogonality penalties against previously computed eigenstates, which constrains each variational search to the subspace orthogonal to lower eigenvectors~\cite{higgott_variational_2019}. This sequential structure makes VQD a natural starting point for a qumode-based excited-state framework, since it requires no modification to the underlying qumode ansatz beyond the addition of computing overlap penalties via quantum or classical methods. We thus introduce the qumode-based variational quantum deflation algorithm (QumVQD), extending the qumode VQE framework of Dutta et al.~\cite{dutta_simulating_2025} to excited states and establishing a framework generalizable to other key quantum chemistry problems.

Beyond electronic structure, the calculation of molecular vibrational energies represents another crucial application domain where bosonic quantum devices offer significant advantages. Despite the importance of vibrational structure calculations, they remain computationally demanding due to the exponential scaling of anharmonic vibrational Hamiltonians with the number of degrees of freedom. Classical computational methods struggle with strongly anharmonic systems and high-dimensional potential energy surfaces. Qubit-based quantum approaches to vibrational structure, including unitary vibrational coupled cluster (UVCC)~\cite{mcardle_digital_2019}, compact heuristic circuit (CHC) ansatzes~\cite{ollitrault_hardware_2020}, and vibrational ADAPT-VQE~\cite{10.1063/5.0191074}, have been benchmarked with small molecules~\cite{r_quantum_2025}, but require costly boson-to-qubit mappings that necessitate truncating the bosonic Fock space and introduce substantial overhead in mapping bosonic unitary transforms to qubit gates~\cite{sawaya_resource-efficient_2020, malpathak_simulating_2025}. The natural representation of vibrational modes as harmonic oscillators makes bosonic quantum devices particularly well-suited for these problems, obviating the need for costly mapping procedures. Recent work has demonstrated the feasibility of digital quantum simulation of vibrational dynamics on bosonic devices using Hamiltonian fragmentation techniques based on the Cartan subalgebra approach~\cite{PRXQuantum.2.040320, malpathak_simulating_2025}, where anharmonic vibrational Hamiltonians are decomposed into solvable fragments that can be efficiently diagonalized using Bogoliubov transforms. By combining this fragmentation technique with our QumVQD framework, we calculate excited-state and ground-state vibrational eigenenergies with minimal quantum circuit depth, thereby improving upon existing qubit-based vibrational structure approaches.

While this manuscript was in preparation, Dutta et al. demonstrated excited-state calculations on bosonic processors using qumode subspace variational quantum eigensolver (QSS-VQE)\cite{dutta_qumode-based_2026}, a parallel subspace-search approach. The present work explores the complementary qumode-based variational quantum deflation framework, termed QumVQD. Both QSS-VQE and QumVQD are excited-state extensions of VQE adapted for qumode-based processors. Beyond the difference in the algorithms used, the present work explores several further directions. First, we incorporate a Hamming-weight filtering technique that enforces particle number conservation in the bosonic VQD setting, reducing the dimension of the effective Hamiltonian and eliminating spurious eigenenergies from the variational search. Second, we incorporate a vibrational Hamiltonian fragmentation approach~\cite{malpathak_simulating_2025} which allows us to calculate vibrational eigenenergies with minimal circuit depth. 
These techniques not only reduce the quantum resource needs for QumVQD, but can potentially also be applied to other qumode-based algorithms for quantum chemistry for further resource reduction.
Further discussion of the relationship between these approaches is provided in the Discussion section.

\section{Particle Number Enforcement} 

\subsection{Motivation and Relation to Prior Work}

Exploiting symmetries to reduce quantum resource requirements is well established in qubit-based quantum computing for quantum chemistry applications. Under the Jordan--Wigner mapping, fixing the total electron count is equivalent to fixing the number of qubits in the $\ket{1}$ state, so that only the $\binom{M}{n_e}$ particle-conserving configurations out of $2^M$ total basis states are physically relevant~\cite{gard_efficient_2020, bravyi_tapering_2017, shee_qubit-efficient_2022}, where $M$ is the total number of spin orbitals, and $n_e$ is the total number of electrons. This observation has motivated symmetry-preserving ansatz circuits that confine the variational search to the correct particle number sector~\cite{gard_efficient_2020}, qubit-efficient encodings that require only $\mathcal{O}(n_e \log_2 M)$ qubits~\cite{shee_qubit-efficient_2022}, and qubit tapering techniques that eliminate redundant degrees of freedom~\cite{bravyi_tapering_2017}. Within VQD specifically, charge-preserving extensions have incorporated fixed particle number into the deflation framework on qubit platforms~\cite{chandani_efficient_2024}.

Here, we adapt this principle to reduce quantum resource requirements in our development of bosonic VQD for molecular excited states. In QumVQD, we make use of the Fock basis encoding of Dutta et al.~\cite{dutta_simulating_2025}. In this encoding, each computational basis state is labeled by a composite Fock index whose binary Hamming weight~\cite{cusick_chapter_2009} equals the electron count under the Jordan--Wigner mapping. Here, the Hamming weight of a number is defined as the count of bits that are ``$1$'' in the binary representation of that number. Restricting to states of fixed Hamming weight, therefore, enforces particle number conservation directly in the qumode register. Although the underlying symmetry enforcement concept is shared with the above qubit methods, the application of Hamming-weight filtering 
for qumode-based variational quantum algorithms has not been previously explored to the best of our knowledge. We provide an analysis of the resulting dimension reduction below.

\subsection{Mathematical Framework}

In the Fock basis representation used by Dutta et al.,~\cite{dutta_simulating_2025} each basis state can be labeled by its Fock index. The total number of 1's (the Hamming weight) in the binary representation of the Fock index equals the number of electrons. By restricting the Hamiltonian to only couple states with identical Hamming weights, particle number conservation is ensured. 

For $M$ total spin orbitals encoded across the qumode register (where $M = N \log_2 d$ for $N$ qumodes of dimension $d$), the unrestricted Hilbert space has dimension $2^M$. Restricting to a fixed electron number $n_e$ reduces the dimension to the combinatorial expression

\begin{equation}
    \mathcal{D}(M, n_e) = \binom{M}{n_e}.
\end{equation}

The resulting Hamiltonian compression depends critically on the filling fraction $n_e / M$. In the dilute regime ($n_e$ fixed, $M \to \infty$), $\binom{M}{n_e} \sim M^{n_e}/n_e!$ is polynomial in $M$, yielding an exponential compression. At half-filling ($n_e = M/2$), Stirling's approximation gives $\binom{M}{M/2} \approx 2^M / \sqrt{\pi M/2}$ as shown in Appendix A, so the reduction is only $\mathcal{O}(\sqrt{M})$. Many quantum chemistry applications fall between these extremes.

Table~\ref{tab:compression} shows exact dimensions for representative molecular systems. Here, $R$ is the ratio between the Hilbert space dimension before particle number enforcement and the corresponding dimension after particle number enforcement. For the small molecules studied in this work (H$_2$, H$_4$, LiH with the STO-3G~\cite{hehre_selfconsistent_1969, collins_selfconsistent_1976} basis set), the $R$ values are modest (from 3 to 8) because these systems are near half-filling. The benefit becomes substantial for larger systems: LiH with the 6-31G~\cite{hehre_selfconsistent_1972} basis set ($n_e/M = 0.18$) achieves a compression ratio $R \approx 573$.

\begin{table}[t]
\centering
\caption{Hilbert space dimensions before and after particle number enforcement 
for representative molecular systems. }
\label{tab:compression}
\renewcommand{\arraystretch}{1.5}

\begin{tabular}{l l c c c r r r c}

\hline
System & Basis & $M$ & $n_e$ & $n_e/M$ & $2^M$ & $\binom{M}{n_e}$ & $R$ & Qumodes ($d{=}16$) \\
\hline
\ce{H2}           & STO-3G & 4  & 2 & 0.50 & 16              & 6     & 2.7  & 1 \\
Linear \ce{H4}    & STO-3G & 8  & 4 & 0.50 & 256             & 70    & 3.7  & 2 \\
\ce{LiH}          & STO-3G & 12 & 4 & 0.33 & 4{,}096         & 495   & 8.3  & 3 \\
\ce{LiH}          & 6-31G  & 22 & 4 & 0.18 & $4.2\times10^6$ & 7{,}315 & 573 & 4 \\
\hline
\end{tabular}
\end{table}

This approach has multiple key benefits. First, it eliminates spurious (non-physical) eigenstates, reducing the number of VQD iterations required to find all physical eigenenergies. Second, the number of qumodes required to store the restricted Hamiltonian is

\begin{equation}
    N_q = \left\lceil \log_d \binom{M}{n_e} \right\rceil
\end{equation}

providing an effective reduction in the number of qumodes relative to the unrestricted encoding.

Importantly, the subspace-restricted Hamiltonians from Hamming-weight filtering cannot directly be applied to physical qumode hardware unless they can be written in terms of bosonic gates. We propose several methods that can be utilized to address this problem in Appendix B. The approach used in this work is to calculate the expectation value of the subspace-restricted Hamiltonian classically as we emulate the qumode-based hardware classically.

\section{Qumode VQD} 

\begin{figure}[H]
    \centering
    \includegraphics[width=1\linewidth]{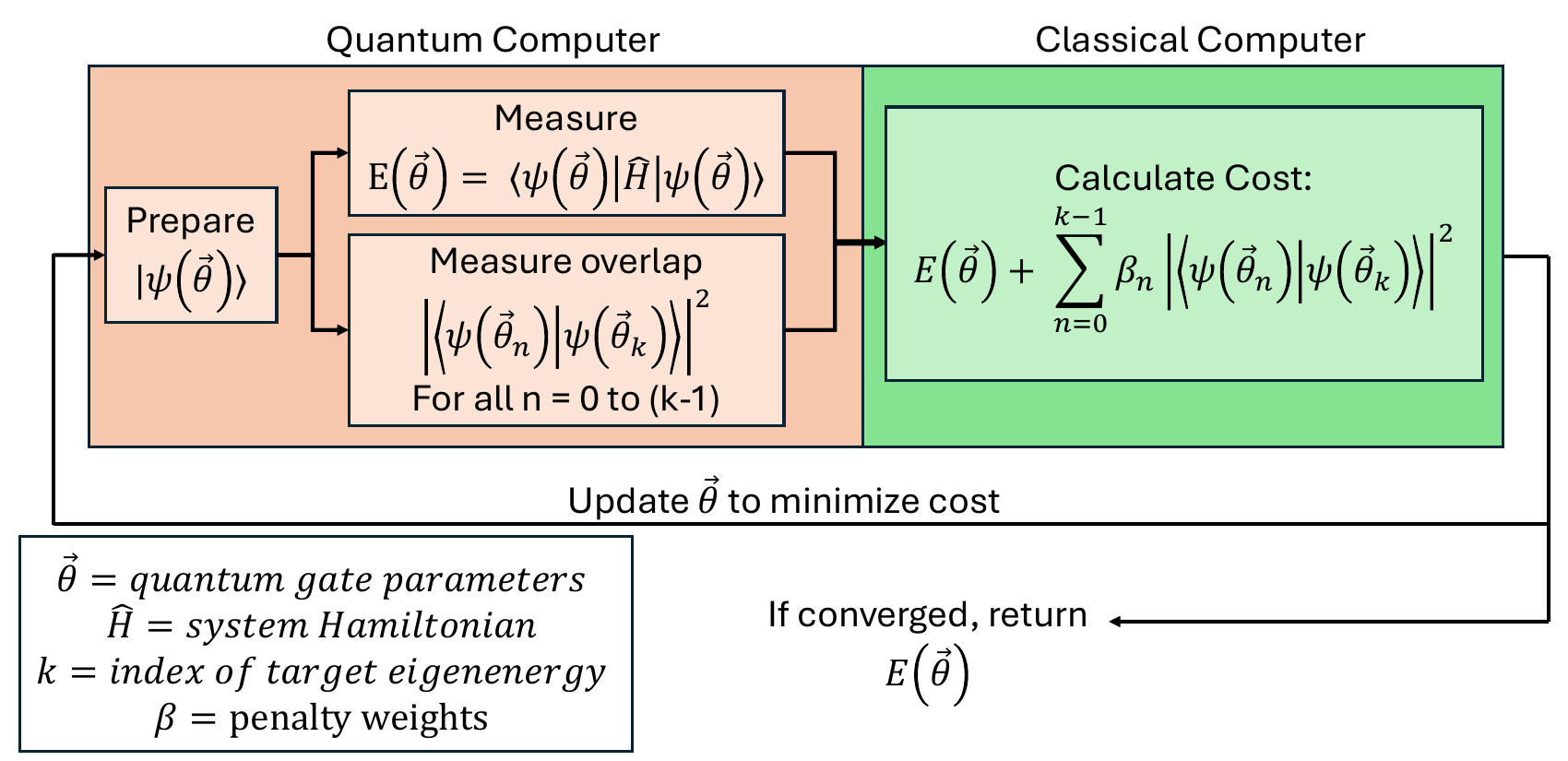}
    \caption{The VQD pipeline. To calculate the energy of the k’th lowest energy eigenstate of a Hamiltonian using VQD, find the state vector with the lowest energy under the constraint that it is orthogonal to all k-1 eigenvectors with lower energy.}
    \label{fig:vqd}
\end{figure}

We extend the VQE ansatz presented by Dutta et al.~\cite{dutta_simulating_2025} for the electronic ground state to compute the electronic excited-state energies of a given problem Hamiltonian using the gate set defined in Appendix C. We implement an extension of VQE known as variational quantum deflation (VQD), which adds a penalty term to VQE’s cost function, penalizing overlap (defined in Figure 1) with previously calculated eigenstates. This orthogonality constraint ensures that the calculated energies are true eigenenergies of the problem Hamiltonian. This constraint can be implemented in multiple ways, either on the classical optimizer or the quantum processor itself~\cite{higgott_variational_2019}. Here, the orthogonality constraint was enforced by computing state overlaps classically and augmenting the cost function with corresponding penalty terms. For each previously optimized state n, the overlap with the current trial wavefunction was weighted by a parameter $\beta_n$ and added to the energy expectation value. The workflow of the VQD pipeline is shown in Figure~\ref{fig:vqd}. 

In our work, QuTiP~\cite{johansson_qutip_2012, johansson_qutip_2013, lambert2025qutip5quantumtoolbox}, NumPy~\cite{2020NumPy-Array}, and TensorFlow~\cite{Abadi_TensorFlow_Largescale_machine_2015, dillon2017tensorflowdistributions} were used for state vector simulations of the qumode device, OpenFermion~\cite{McClean_OpenFermion_The_Electronic_2020} and OpenFermion-PySCF~\cite{McClean_OpenFermion_The_Electronic_2020,  https://doi.org/10.1002/wcms.1340} were used to supply electronic structure Hamiltonians and to map them to qubit Hamiltonians, and PySCF~\cite{sun_recent_2020} was used for full configuration interaction (FCI)~\cite{doi:https://doi.org/10.1002/9781119019572.ch11, knowles_new_1984} calculations. Following the convention in Dutta et al.~\cite{dutta_simulating_2025}, in our framework, QumVQD, we define a parameter $D$ to indicate the number of concatenated SNAP and displacement gate pairs, which quantifies the depth of an ansatz. In cases where the quantum circuit uses multiple qumodes, each pair of SNAP and displacement gates is accompanied by a set of beam splitter (BS) gates with all-to-all connectivity. Unless otherwise specified, the Fock cutoff for each qubit is taken to be 16 as demonstrated experimentally~\cite{wang_efficient_2020}. Throughout this paper, we define chemical accuracy as 1 kcal/mol~\cite{r_quantum_2025} and spectroscopic accuracy as 1 $\text{cm}^{-1}$~\cite{puzzarini_accuracy_2019}.

\subsection{Electronic Structure}

\begin{figure*}[t]
    \includegraphics[width=\linewidth]{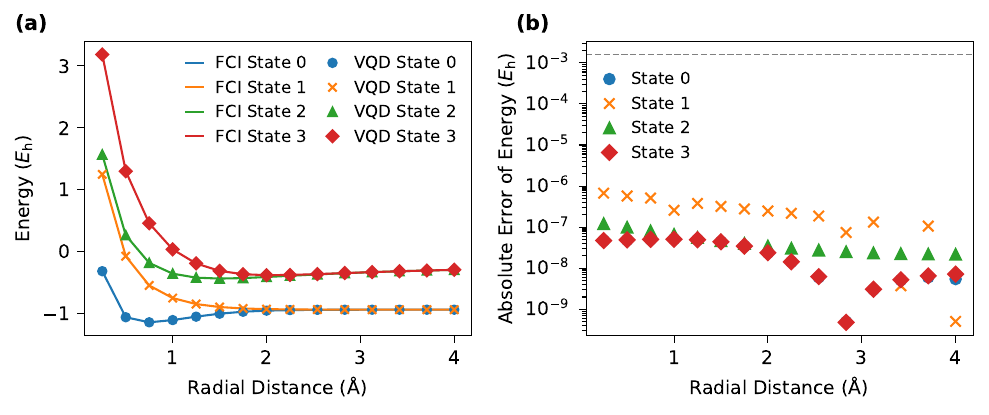}
    \caption{QumVQD results for the electronic eigenenergies of \ce{H2}. (a) Potential energy surfaces for the ground and excited states of \ce{H2} calculated with QumVQD compared to FCI. (b) Absolute error of the \ce{H2} PESs from QumVQD using FCI as the reference. The horizontal grey dashed line indicates chemical accuracy.}
    \label{fig:h2_pes}
\end{figure*}

With particle number enforcement, we compare the energies resulting from QumVQD with the energies from FCI (Figure~\ref{fig:h2_pes}) for the potential energy surfaces (PESs) of \ce{H2}. For this system, the ansatz depth $D$ was between 4 and 7, with higher depths used in some cases to help reduce the number of failed convergences during optimization. The specific values can be found in the code associated with this work~\cite{jost_2026_20479296}. All overlap penalty weights $\beta_n$ were set to 5.0 and only one qumode with Fock cutoff of 6 was required. Two filters were applied to the energies calculated by QumVQD. First, we filtered out any states with an overlap above $10^{-10}$ with the lower energy eigenstates. Second, redundant copies of the same eigenenergy were removed. All remaining energies calculated by the simulated QumVQD circuit were well within chemical accuracy. Note that the 1st excited state of H$_2$, $^3\Sigma_u^+$, results in three degenerate eigenstates with VQD~\cite{higgott_variational_2019}, and we plot only one representative for visual clarity.

\begin{figure*}[t]
    \centering
    \includegraphics[width=1\linewidth]{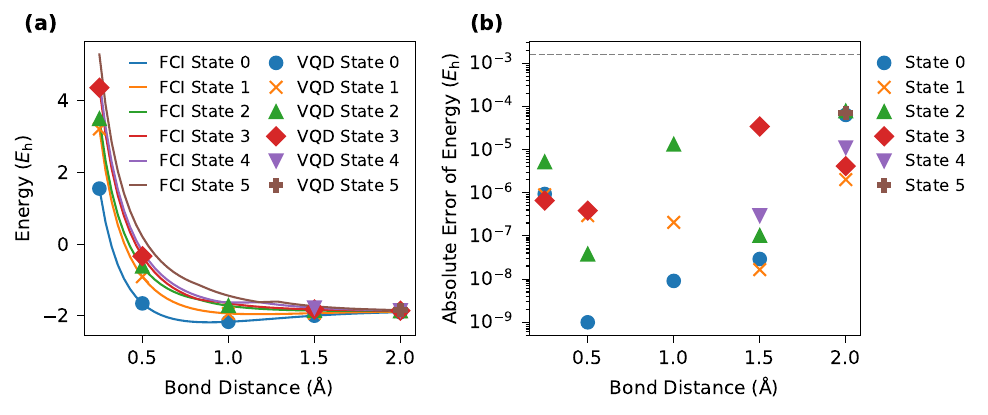}
    \caption{QumVQD results for the electronic eigenenergies of linear \ce{H4}. (a) Potential energy surfaces for the lowest five eigenstates of \ce{H4} calculated with QumVQD compared to the corresponding values calculated with FCI. (b) Absolute error of the \ce{H4} PESs from QumVQD using FCI as the reference. The horizontal grey dashed line indicates chemical accuracy. Not all five energies were calculated for each bond distance. Note that both QumVQD and FCI curves represent the n-th lowest eigenenergy at each bond distance rather than a continuously tracked eigenstate, so apparent kinks occur where two states cross and the energy ordering switches.}
    \label{fig:h4_sliced_pes}
\end{figure*}

For linear \ce{H4}, we compare the energies resulting from QumVQD using the Hamming-weight filtered Hamiltonian with FCI energies (Figure~\ref{fig:h4_sliced_pes}). The depth $D$ and $\beta_n$ values were again increased in some cases to help reduce the number of failed convergences during optimization. The specific values can be found in the code associated with this work~\cite{jost_2026_20479296}. The results in Figure~\ref{fig:h4_sliced_pes} were collected and filtered from multiple executions of QumVQD to mitigate the impact of failed convergences of the classical optimizer. Such failed convergences can occur, for example, when the optimizer gets trapped in a local minimum. As with \ce{H2}, we filtered out states with low overlap with the lower energy eigenstates and removed redundant copies of the same eigenenergy, though here the overlap threshold was $10^{-7}$. For some distances, not all of the lowest six eigenenergies were found, but this can be remedied by executing QumVQD more times.

\subsection{Vibrational Structure} 

The previous section demonstrates that we have successfully implemented bosonic VQD for excited-state electronic structure calculations. We now apply the methodology to calculating the eigenenergies of molecular vibrations.

Vibrational Hamiltonians can be decomposed into easily solvable fragments as shown in Equation~\ref{eqn:fragments}, where $D_k$ are diagonal~\cite{malpathak_simulating_2025, malpathak_trotter_2026}. This fragmentation framework was originally developed for Hamiltonian simulation on bosonic hardware~\cite{malpathak_simulating_2025} and subsequently extended to qubit-based Trotter simulation~\cite{malpathak_trotter_2026}, in both cases primarily targeting vibrational dynamics. Here, we adapt it to the measurement step of a variational eigenvalue framework on bosonic hardware, enabling direct calculation of vibrational eigenenergies rather than time evolution.

\begin{equation}
    H = \sum_k H_k = \sum_k U_kD_kU_k^\dagger
    \label{eqn:fragments}
\end{equation}

$U_k$ can be applied using qumode gates as shown in Equation~\ref{eqn:U_k gates} with N being the number of qumodes and the parameters $\gamma, \phi, \zeta,$ and $\chi$ calculated during the greedy optimization loop employed by Malpathak et al.~\cite{malpathak_simulating_2025} The corresponding gate definitions are given in Appendix C. 

To measure the energy of a trial state $\ket{\psi}$, one simply applies the gate decomposition of $U_k$ (Equation~\ref{eqn:U_k gates}~\cite{malpathak_simulating_2025}) to $\ket{\psi}$ before performing photon-number measurements to calculate the expectation of $D_k$ as shown in Equation~\ref{eqn:expectation of D_k}. 

\begin{equation}
    U_k = \left( \prod_{p=1}^{N} D_p(\gamma_p) \right) 
    \left( \prod_{p>q=1}^{N} BS_{pq} \left( 2\phi_{pq}, \frac{\pi}{2} \right) \right)   
    \times \left( \prod_{p=1}^{N} S_p(\zeta_p, 0) \right) 
    \left( \prod_{p>q=1}^{N} BS_{pq} \left( 2\chi_{pq}, \frac{\pi}{2} \right) \right)
    \label{eqn:U_k gates}
\end{equation}

\begin{equation}
    \braket{\psi|H_k|\psi} = \braket{\psi|U_kD_kU_k^\dagger|\psi}
    = \braket{\psi'|D_k|\psi'}
    = \braket{D_k}
    \label{eqn:expectation of D_k}
\end{equation}

\begin{figure}[t]
    \includegraphics[width=\linewidth]{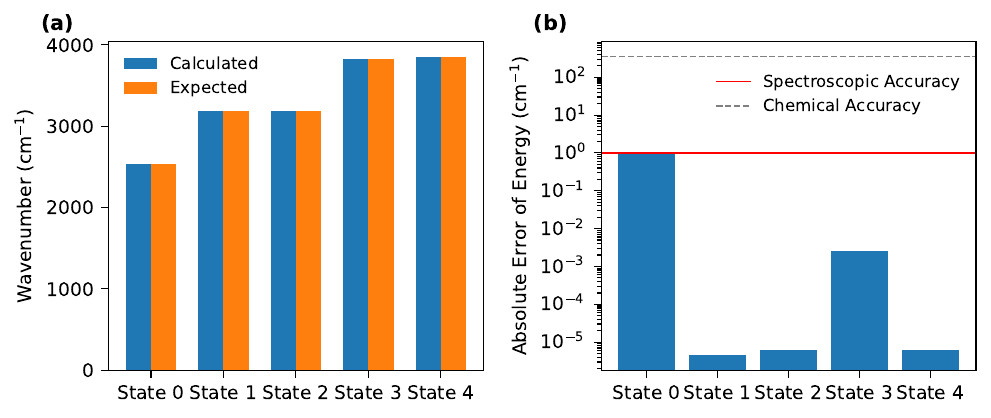}
        \caption{QumVQD results for the vibrational eigenenergies of \ce{CO2}. (a) Comparison of the five lowest vibrational eigenenergies of \ce{CO2} using QumVQD (Calculated) and direct diagonalization (Expected). (b) Absolute error in the five lowest vibrational eigenenergies of \ce{CO2} from QumVQD compared to direct diagonalization.}
    \label{fig:co2_vqd}
\end{figure}

\begin{figure}[t]
        \includegraphics[width=\linewidth]{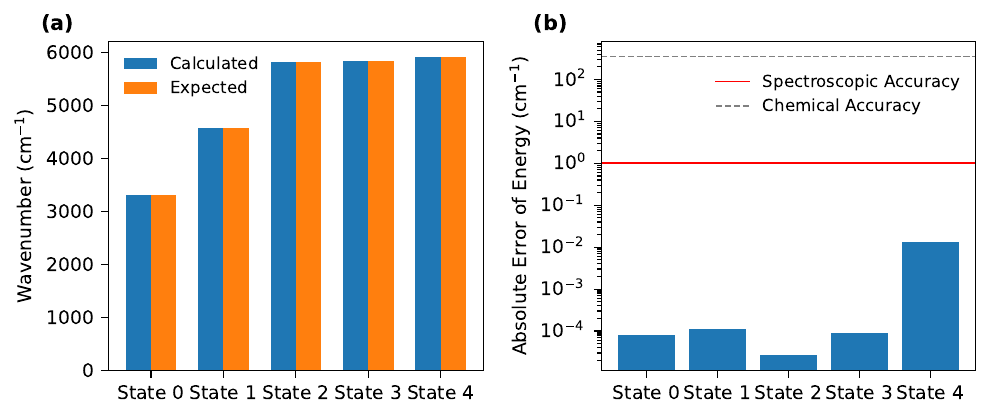}
        \caption{QumVQD results for the vibrational eigenenergies of \ce{H2S}. (a) Comparison of the five lowest vibrational eigenenergies of \ce{H2S} using QumVQD (Calculated) and direct diagonalization (Expected). (b) Absolute error in the five lowest vibrational eigenenergies of \ce{H2S} from QumVQD compared to direct diagonalization.}
    \label{fig:h2s_vqd}
\end{figure}

We simulate a QumVQD circuit for each vibrational fragment with $D = 10$ for \ce{CO2} and $D = 25$ for \ce{H2S}. For our simulation, we apply the vibrational fragments directly to a single simulated qumode, with Fock cutoff of 64 for \ce{H2S} and 256 for \ce{CO2}, to find the eigenenergies of the original vibrational Hamiltonian. This simplifies the simulation process for implementing Equations~\ref{eqn:U_k gates} and~\ref{eqn:expectation of D_k}. 
The energies resulting from our simulated VQD circuit are compared with those found through direct diagonalization of the original vibrational Hamiltonians in Figures~\ref{fig:co2_vqd} and~\ref{fig:h2s_vqd} for \ce{CO2} and \ce{H2S}, respectively.

The sparse matrix representations of the fragment operators $U_k D_k U_k^\dagger$ were constructed using \texttt{openfermion.linalg.boson\_operator\_sparse}~\cite{McClean_OpenFermion_The_Electronic_2020} with a Fock-space truncation of 4. We verified that diagonalizing the fragments at this truncation reproduces the eigenenergies of the unfragmented Hamiltonian at the same truncation to within $10^{-9}$ cm$^{-1}$, confirming that the fragmentation procedure preserved the original eigenenergies. The QumVQD eigenenergies reported below are therefore compared against direct diagonalization at the same truncation, isolating the algorithmic performance from truncation effects.

To our knowledge, the fragmentation approach of Malpathak et al.~\cite{malpathak_simulating_2025} has not been previously incorporated into such a VQD workflow. This fragmentation approach offers two key advantages for QumVQD. First, it decomposes the vibrational Hamiltonian into fragments that are each easily tractable on qumode-based hardware, helping minimize circuit depth. Second, because the energy expectation value of a trial state under the full Hamiltonian equals the sum of its expectation values under the individual fragments, $\braket{\psi|H|\psi} = \sum_k \braket{\psi|H_k|\psi}$, the measurement step of QumVQD is naturally parallelizable. The same trial state $\ket{\psi}$ can be prepared on separate quantum processors, one per fragment, and each expectation value can be measured simultaneously. These values are then summed classically to obtain the total energy for the classical optimizer. This reduces the circuit depth required for any single quantum processor.

Using Equation~\ref{eqn:U_k gates} to implement the vibrational Hamiltonian requires one displacement gate and one one-mode squeezing gate per qumode, plus $N(N-1)$ beam splitter gates for $N$ qumodes, assuming all-to-all connectivity. Including an ansatz of depth $D$ adds $D$ SNAP and $D$ displacement gates per qumode, along with an additional $DN(N-1)/2$ beam splitter gates. The total non-trivial gate count thus scales as $\mathcal{O}(N)$ in single-qumode gates and $\mathcal{O}(N^2)$ in beam splitter gates per fragment. For the example of \ce{CO2}, which has four modes~\cite{malpathak_simulating_2025}, this would yield 12 beam splitter, 4 displacement, and 4 squeezing gates per fragment. With an ansatz of depth $D=20$, and 4 bosonic fragments~\cite{malpathak_simulating_2025}, the total gate counts would be 168 beam splitter, 96 displacement, 80 SNAP, and 16 squeezing gates in total. In contrast, qubit-based unitary vibrational coupled cluster (UVCC)~\cite{mcardle_digital_2019} and compact heuristic circuit (CHC)~\cite{ollitrault_hardware_2020} approaches to the \ce{CO2} vibrational Hamiltonian with 4 modals have been reported to require over 2{,}000 and over 300 two-qubit controlled-X (CX) gates, respectively~\cite{r_quantum_2025}. Moreover, the number of CX gates required by UVCC and CHC grows rapidly with the number of modals per vibrational mode~\cite{ollitrault_hardware_2020, r_quantum_2025}, since each additional modal requires additional qubits and entangling gates. In the qumode encoding, the number of modals corresponds to the Fock-space truncation of each qumode, so the QumVQD gate count depends only on the number of qumodes and the ansatz depth, and is therefore independent of this truncation, provided the Fock cutoff is large enough to capture the maximum occupation of each mode.

\section{Discussion} 

Using QumVQD, the electronic structure calculations for \ce{H2} and \ce{H4} achieved high accuracy, with all computed excited-state energies matching FCI benchmarks to within $10^{-3}$ hartree, well below the chemical accuracy threshold of 1 kcal/mol 
(Figure~\ref{fig:h2_pes}). This precision was maintained across multiple excited states and various molecular geometries. 
The ability of bosonic VQD to reproduce FCI energies across these benchmarks demonstrates that the qumode VQD framework converges to the correct excited-state spectrum within or near chemical accuracy on small electronic systems. Likewise, the vibrational energy errors from QumVQD are at or within the spectroscopic accuracy threshold of 1 cm$^{-1}$.
 
The bosonic gate counts for these calculations are given in Table~\ref{tab:gate_counts}. As discussed in the previous section, the gate count for QumVQD to describe vibrational excited states is significantly less than that from qubit-based algorithms.  
Our results establish that bosonic quantum devices may be able to calculate electronic and vibrational energies with high accuracy and few quantum resources, particularly when combined with symmetry enforcement and Hamiltonian fragmentation strategies.

\begin{table}[h]
\centering
\caption{Gate count scaling for QumVQD circuit components by Hamiltonian type and ansatz layer. N denotes the number of qumodes, $D$ the ansatz depth, $N_H$ the number of Pauli words in the qubit-mapped electronic Hamiltonian, and $N_d$ the number of SNAP and displacement gates used to approximate a Pauli word~\cite{dutta_simulating_2025}. In this work, $N_H=\mathcal{O}(M^4)$ for $M$ spin orbitals, as we use the Jordan--Wigner Transformation~\cite{dutta_simulating_2025}, although more efficient mappings are available~\cite{PhysRevX.8.011044}.}
\label{tab:gate_counts}
\renewcommand{\arraystretch}{1.5}

\begin{tabular}{llr}

\hline
\textbf{Component} & \textbf{Gate Type} & \textbf{Count} \\
\hline
Ansatz (count per layer)       & SNAP          & 1 per qumode \\
                        & Displacement  & 1 per qumode \\
                        & Beam Splitter & $N(N-1)/2$   \\
\hline
Electronic Hamiltonian  & SNAP and Displacement & $\mathcal{O}(N_H N_d)$ \\
\hline
Vibrational Hamiltonian (count per fragment) & Displacement  & $N$      \\
                        & Squeezing     & $N$          \\
                        & Beam Splitter & $N(N-1)$     \\
\hline
\end{tabular}
\end{table}

This work provides several advances regarding quantum chemistry on bosonic devices compared to existing work, such as QSS-VQE~\cite{dutta_qumode-based_2026}. First, we apply bosonic VQD to vibrational structure calculations using the Hamiltonian fragmentation approach of Malpathak et al.\cite{malpathak_simulating_2025}, demonstrating that the combination of native bosonic representation with fragmentation strategies yields dramatic reductions in gate counts compared to qubit-based approaches. 
Second, we demonstrate the use of a Hamming-weight-based symmetry enforcement technique that achieves a dimension reduction scaling with system size, complementing the two-qubit reduction provided by Qiskit's ParityMapper~\cite{developers_qiskit_2023, wang_characterizing_2025} used in the work of Dutta et al.~\cite{dutta_qumode-based_2026}.
QSS-VQE prepares $k$ orthogonal Fock states at each optimization iteration to access $k$ eigenstates in parallel, while QumVQD prepares a single vacuum state and accesses eigenstates sequentially via deflation. The parallel approach may offer reduced wall-clock cost per eigenstate when many eigenstates are needed simultaneously, while the sequential approach reduces per-iteration state preparation overhead. QumVQD additionally requires the computation of $k-1$ state overlaps at each iteration when targeting the $k$-th eigenstate, an overhead avoided by QSS-VQE's subspace structure, which preserves orthogonality by construction. These overlaps can be computed either on the quantum processor via SWAP or Hadamard tests, which increases the circuit depth, or classically from measured state vectors, which shifts the cost to classical post-processing~\cite{higgott_variational_2019} and requires many repeated measurements. On the other hand, QumVQD allows the target number of eigenstates to be extended after an initial calculation: an additional eigenstate is obtained by appending one deflation term and running a single new optimization, without revisiting previously found states as QSS-VQE would require. We leave the quantitative comparison of the resource requirements, accuracy, and noise susceptibility of these complementary approaches for excited-state calculations on bosonic processors for future work. 

\section*{Conclusions}
We have developed QumVQD, a framework for computing excited-state
energies on qumode-based processors, applicable to both electronic and vibrational Hamiltonians within a single variational scheme. For electronic structure, particle number conservation is enforced through Hamming-weight filtering of the Fock basis, a symmetry-reduction strategy that maps naturally onto the qumode encoding and reduces the relevant Hilbert space dimension from $2^M$ to $\binom{M}{n_e}$ for $M$ spin orbitals and $n_e$ electrons. This reduction is most pronounced away from half-filling, where the restricted dimension $\binom{M}{n_e}$ scales polynomially in $M$ rather than exponentially as $2^M$ does. QumVQD reproduces the FCI potential energy surfaces of \ce{H2} to within chemical accuracy across the bond dissociation curve. For linear \ce{H4}, multiple QumVQD runs combined with overlap and uniqueness-based filtering recover the lowest six excited-state energies to chemical accuracy for most geometries, with the requirement of multiple runs reflecting the non-convex structure of the deflated cost function.

For vibrational structure, pairing QumVQD with the Bogoliubov-transformation-based fragmentation of Malpathak et al.~\cite{malpathak_simulating_2025} yields per-fragment circuits requiring $N$ displacement, $N$ squeezing, and $N(N-1)$ beam splitter gates to implement each fragment unitary, together with $D$ SNAP, $D$ displacement, and $DN(N-1)/2$ beam splitter gates from the ansatz layers, for $N$ qumodes at depth $D$. For \ce{CO2} at the system sizes considered, this yields total non-trivial gate counts substantially below those reported for qubit-based unitary vibrational coupled cluster and compact heuristic circuit approaches to the corresponding unfragmented Hamiltonians, while reaching spectroscopic accuracy on the five lowest vibrational eigenstates of each molecule. The reduction reflects the combined effect of working in the native bosonic representation, which eliminates boson-to-qubit mapping overhead, and of distributing the Hamiltonian across fragments that can be evaluated independently against a shared trial state. With fewer gates, less error accumulation is expected on noisy devices. Detailed analysis of whether or how this reduction in gate count improves the error rates of these eigenenergy calculations when using qumodes is left to future work.

Taken together, these results establish QumVQD as a concrete route to excited-state quantum chemistry on bosonic hardware, with the application to vibrational excited states showing the clearest resource advantage over qubit-based alternatives.

\appendix

\section{Appendix A: Stirling's Approximation for $\binom{M}{M/2}$}
\label{app:stirling}
\renewcommand{\theequation}{A\arabic{equation}}
\setcounter{equation}{0}

Using Stirling's approximation, $n! \sim \sqrt{2\pi n}(n/e)^{n}$, the central binomial coefficient becomes

\begin{equation}
\binom{M}{M/2} = \frac{M!}{\left[(M/2)!\right]^2} \sim  \frac{\sqrt{2\pi M}(\frac{M}{e})^{M}}{\pi M\,(\frac{M}{2e})^{M}} = \sqrt{\frac{2}{\pi M}}\,2^{M}
\end{equation}

for even $M$. 

\section{Appendix B: Hardware implementation strategies for subspace-restricted Hamiltonians from Hamming-weight filtering}
\label{app:hardware implementation}

One method is to classically approximate the subspace-restricted Hamiltonian with a series of SNAP, beam splitter, and displacement gates. Since these gates form a universal gate set~\cite{zhang_energy-dependent_2024, you_crosstalk-robust_2024}, the subspace-restricted Hamiltonian can, in principle, be reproduced to arbitrary accuracy by optimizing the parameters of a sufficiently deep sequence of such gates against the target operator. However, for large enough molecules, this classical approximation may prove challenging.

A second method is to apply our Hamming-weight filtering directly to each of the original SNAP and displacement gates that represented a given molecular Hamiltonian. From there, each reduced SNAP and displacement gate must be mapped to new SNAP and displacement gates that can physically act on the reduced Hilbert space. In the single-qumode case, reduced SNAP gates are straightforward to map to physically implementable SNAP gates on the reduced Hilbert space, as the SNAP gate is diagonal in the Fock basis. Namely, for a SNAP gate parametrized by the vector $\theta$, when the indices of the problem Hamiltonian not in the set $\{i\}$ are filtered out, the new SNAP gate is simply a SNAP gate parametrized by $\theta[j]$ where $j \in \{i\}$. Reduced displacement gates are more involved, as the displacement operator is not diagonal in the Fock basis and its action couples Fock states outside the retained subspace. Nevertheless, the resulting reduced operator can be recompiled into a sequence of SNAP and displacement gates acting on the reduced Hilbert space. This single-qumode strategy incurs some overhead in gate count for the expectation measurement portion of the circuit. Extending this approach to the multi-qumode setting is less straightforward, as the entangling structure between qumodes complicates the recompilation of reduced operators, and we leave a systematic treatment of this case to future work.

A third method is to measure the Fock state of the qumode after applying the ansatz to the vacuum state and then reconstruct the wavefunction classically. From there, the expectation value of the subspace-restricted Hamiltonian can be calculated classically. While this shifts a portion of the workload to a classical post-processing step, the dimensions of the state vector to be measured and the matrix to be stored and multiplied are already substantially reduced by the Hamming-weight filtering, mitigating the classical cost. This approach is the procedure used in our numerical simulations as it is simple to calculate and pairs naturally with the classical overlap measurement used in our deflation step. 

A fourth method is to use qubit-based particle-number enforcement techniques such as the Qubit-Efficient Encoding scheme~\cite{shee_qubit-efficient_2022} (QEE) before mapping the qubit Hamiltonian to a bosonic one. 
The main drawback of this route, relative to a direct Jordan--Wigner mapping followed by bosonic compilation, is an increase in the number of terms in the qubit Hamiltonian, which has an upper bound of $\mathcal{O}(\frac{M^{2n_e+1}}{(n_e-1)!n_e!})$ in the QEE scheme compared to $\mathcal{O}(M^4)$ for Jordan--Wigner~\cite{shee_qubit-efficient_2022}. Because each Pauli word maps to a fixed-depth bosonic gate sequence under the mapping of Dutta et al.~\cite{dutta_simulating_2025}, this term-count increase translates directly into a proportional increase in the number of controlled-unitary measurements required for the expectation value.

Each of the four methods above presents a distinct trade-off between classical and quantum resources, and the optimal choice will, in general, depend on the system size, available hardware, and target accuracy.

\section{Appendix C: Bosonic Gates}
\label{app:gates}

\renewcommand{\theequation}{C\arabic{equation}}
\setcounter{equation}{0}

This work uses four kinds of bosonic gates. The first is the selective number-dependent arbitrary phase (SNAP) gate~\cite{heeres_cavity_2015}

\begin{equation}
    SNAP(\theta) = \sum^{L-1}_{n=0} e^{i\theta_n}\ket{n}\bra{n}
\end{equation}

where L is the Fock cutoff for each qumode. The second is the displacement gate~\cite{liu_hybrid_2026} 

\begin{equation}
    D(\alpha) = e^{\alpha \hat{b}^\dagger - \alpha^*\hat{b}}
\end{equation}

where $\hat{b}^\dagger$ and $\hat{b}$ are the bosonic raising and lowering operators respectively. Together, the SNAP and displacement gates form a universal single qumode unitary~\cite{krastanov_universal_2015}. The third is the two-qumode entangling beam splitter (BS) gate~\cite{liu_hybrid_2026}

\begin{equation}
    BS_{j,k}(\beta, \phi) = e^{i\beta/2 (e^{i\phi}\hat{b}^\dagger_j \hat{b}_k + e^{-i\phi}\hat{b}^\dagger_k \hat{b}_j)}
\end{equation}

which, together with a universal single-qumode gate set, forms a universal multimode gate set~\cite{zhang_energy-dependent_2024, you_crosstalk-robust_2024}. Additionally, we utilize the one-mode squeezing gate~\cite{braunstein_quantum_2005} 

\begin{equation}
        S(\gamma) = e^{\frac{1}{2}\left(\gamma \hat{b}^\dagger{}^2 - \gamma \hat{b}^2\right)}
\end{equation}

for vibrational structure calculations.

\section*{Author contributions}

Marlon F. Jost: Conceptualization (equal); Investigation (lead); Methodology (equal); Software (lead); Writing -- original draft (lead); Funding acquisition (supporting). 
Sijia S. Dong: Conceptualization (equal); Methodology (equal); Project administration (lead); Resources (lead); Writing -- review \& editing (lead); Funding acquisition (lead); Supervision (lead).

\section*{Conflicts of interest}
There are no conflicts to declare.

\section*{Data availability}
The source code, simulation notebooks, and data needed to reproduce all figures in this work are available at \href{https://github.com/sdonglab/QumVQD}{github.com/sdonglab/QumVQD} and archived at \href{https://doi.org/10.5281/zenodo.20479296}{zenodo.20479296}~\cite{jost_2026_20479296}.

\section*{Acknowledgement}

This work was supported by the Arnold and Mabel Beckman Foundation through the Beckman Scholars Program 2025 grant (Crossref ID: dx.doi.org/10.13039/100000997), by the Northeastern University startup fund, and by the U.S. Department of Energy, Office of Science, Office of Advanced Scientific Computing Research, under Award Number DE-SC0024216. The authors thank Artur F. Izmaylov, Shreyas Malpathak, and Sangeeth Das Kallullathil for their fragmentation code and for helpful discussions.

\bibliography{References}

\end{document}